\documentclass[10pt,journal,a4paper]{IEEEtran}
\usepackage[T1]{fontenc}
\usepackage[latin1]{inputenc}
\usepackage{graphicx}
\usepackage{theorem}
\usepackage{amsmath}
\usepackage{amsfonts}
\usepackage[caption=false,font=footnotesize]{subfig}
\usepackage{acronym}

\acrodef{QC}{quasi-cyclic}
\acrodef{QC-LDPC}{quasi-cyclic low-density parity-check}
\acrodef{LDPC}{low-density parity-check}
\acrodef{LDPCC}{low-density parity-check convolutional}
\acrodef{AC-LDPC}{array convolutional low-density parity-check}
\acrodef{PDC-LDPC}{progressive difference convolutional low-density parity-check}
\acrodef{AWGN}{additive white Gaussian noise}
\acrodef{BER}{bit error rate}
\acrodef{FER}{frame error rate}
\acrodef{TUB}{truncated union bound}
\acrodefplural{TUB}[TUBs]{truncated union bounds}
\acrodef{BPSK}{binary phase shift keying}
\acrodef{SPA-LLR}{sum-product algorithm with log-likelihood ratios}
\acrodef{RTI}{regular time-invariant}
\acrodef{RTI-LDPCC}{regular time-invariant low-density parity-check convolutional}

\makeatletter

{\theorembodyfont{\rmfamily}
   }
{\theorembodyfont{\rmfamily}
   }
{\theorembodyfont{\rmfamily}
   }
{\theorembodyfont{\rmfamily}
   }
{\theorembodyfont{\rmfamily}
   }
{\theorembodyfont{\rmfamily}
   \newtheorem{Exa}{{\textbf Example}}[section]}
 
\def\HH{\mathbf{H}}
\def\0{\mathbf{0}}
\def\Hconv{\HH_\mathrm{conv}}

\makeatother
\begin{document}

\title{{\huge Array Convolutional Low-Density Parity-Check Codes}}

\author{Marco Baldi,~\IEEEmembership{Senior~Member,~IEEE}, Giovanni Cancellieri, and Franco Chiaraluce,~\IEEEmembership{Member,~IEEE}

\thanks{Copyright (c) 2013 IEEE. Personal use of this material is permitted. However, permission to use this material for any other purposes must be obtained from the IEEE by sending a request to pubs-permissions@ieee.org.

M. Baldi, G. Cancellieri and F. Chiaraluce are with the Dipartimento di Ingegneria dell'Informazione, Università Politecnica delle Marche, Ancona, Italy (e-mail: \{m.baldi, g.cancellieri, f.chiaraluce\}@univpm.it).

This work was supported in part by the MIUR project ``ESCAPADE'' (Grant RBFR105NLC) under the ``FIRB - Futuro in Ricerca 2010'' funding program.}}

%
%

\maketitle
\begin{abstract}
This paper presents a design technique for obtaining
\ac{RTI-LDPCC} codes with low complexity and good performance.
We start from previous approaches which unwrap a 
\ac{LDPC} block code into an \ac{RTI-LDPCC} code,
and we obtain a new method to design \ac{RTI-LDPCC} codes with better
performance and shorter constraint length.
Differently from previous techniques, we start the design from an array
\ac{LDPC} block code.
We show that, for codes with high rate, a performance gain and a reduction in the
constraint length are achieved with respect to previous 
proposals. Additionally, an increase in the minimum distance 
is observed.
\end{abstract}


\section{Introduction}

\ac{LDPC} codes \cite{Gallager} are the state of the art in forward error correction,
because of their capacity-achieving performance under belief propagation decoding \cite{Richardson2001}.
These decoding algorithms work on the code Tanner graph, corresponding to the parity-check matrix $\HH$, and have a complexity which increases linearly with the code length.
Their best performance is achieved by using irregular codes, with degree distributions optimized through density evolution \cite{Chung2001}.
However, regular codes incur a loss with respect to irregular codes which decreases with increasing code rate, and becomes very small at high code rates, which are of interest in many practical applications.
In addition, the code regularity simplifies the hardware implementation of the belief propagation decoder and allows a scalable design.
Structured \ac{LDPC} codes, like \ac{QC-LDPC} codes \cite{Lin2004Book, Baldi2011a} and \ac{LDPCC} codes \cite{Tanner2004, Felstrom1999, Pusane2011} can also take advantage of simple encoding circuits.

In this paper, we focus on \ac{RTI-LDPCC} codes.
As for \ac{QC-LDPC} codes, \ac{LDPCC} codes can be encoded through simple circuits based on shift registers like that in \cite[Fig. 4]{Felstrom1999}.
Their decoding can be performed by running a belief propagation algorithm on a window sliding over the received sequence \cite{Felstrom1999}.
Both these encoding and decoding methods have a complexity which increases linearly with the code syndrome former constraint length.
Hence, we use the syndrome former constraint length as a measure of the complexity, and we aim at designing codes with small constraint lengths.

A well-known approach for designing time invariant \ac{LDPCC} codes has been proposed by Tanner et al. in \cite{Tanner2004}.
The codes designed in \cite{Tanner2004} are both regular and irregular and usually have large constraint lengths.
Unfortunately, the \ac{RTI-LDPCC} codes designed as in \cite{Tanner2004} often exhibit high error floors at high code rates.
This was already observed in \cite{Baldi2012b}, where another class of \ac{RTI-LDPCC} codes,
named \ac{PDC-LDPC} codes, was presented.
Those codes have a known minimum distance and achieve strong reductions in the constraint length with respect to
the approach in \cite{Tanner2004}. On the other hand, \ac{PDC-LDPC} codes with high rate also exhibit rather high error floors,
and do not outperform those designed following \cite{Tanner2004}.

We propose a method to design \ac{RTI-LDPCC} codes
which represent a further step in this direction.
In fact, the proposed codes, similarly to those in \cite{Baldi2012b}, achieve a smaller constraint length than their 
counterparts designed as in \cite{Tanner2004} but, contrary to \cite{Baldi2012b}, they are also able to achieve 
better performance at high code rates, which are of major interest for this kind of codes.
Differently from previous solutions, the proposed design starts from a special class of \ac{QC-LDPC} codes, named array \ac{LDPC} codes \cite{Fan2000}.
For this reason, we call these codes \ac{AC-LDPC} codes.
The organization of the paper is as follows:
in Section \ref{sec:arrayLDPC}, we remind the main characteristics of array \ac{LDPC} codes;
in Section \ref{sec:ACLDPC}, we define the new class of \ac{AC-LDPC} codes;
in Section \ref{sec:Comparison}, we compare this approach with previously proposed solutions;
in Section \ref{sec:Examples}, we provide some design examples and conclusive remarks.

\section{Array LDPC codes \label{sec:arrayLDPC} }

Array LDPC codes are a special class of \ac{QC-LDPC} codes.
A \ac{QC} code has dimension $k$ and length $n$ which 
are both multiple of a positive integer $q$, i.e., $k = k_0q$ and $n = n_0q$.
Hence, the code redundancy is $r = (n_0 - k_0)q = r_0q$.
In a \ac{QC} code, every cyclic shift of 
$n_0$ positions of a codeword yields another codeword. 
When a \ac{QC} code is also an \ac{LDPC} code, it is called a \ac{QC-LDPC} code.
The parity-check matrix $\mathbf{H}$ of a \ac{QC} code has the following form:

\begin{equation}
\mathbf{H}=\left[\begin{array}{llll}
\mathbf{H}_{0,0}^{c} & \mathbf{H}_{0,1}^{c} & \ldots & \mathbf{H}_{0,n_{0}-1}^{c}\\
\mathbf{H}_{1,0}^{c} & \mathbf{H}_{1,1}^{c} & \ldots & \mathbf{H}_{1,n_{0}-1}^{c}\\
\vdots & \vdots & \ddots & \vdots\\
\mathbf{H}_{r_{0}-1,0}^{c} & \mathbf{H}_{r_{0}-1,1}^{c} & \ldots & \mathbf{H}_{r_{0}-1,n_{0}-1}^{c}\end{array}\right].
\label{eq:H_BlocksOfCirc}
\end{equation}
In \eqref{eq:H_BlocksOfCirc}, each $\mathbf{H}_{i,j}^{c}$ is a $q\times q$ circulant matrix, i.e.,
a square matrix whose $l$-th row, $l = 0,1,2,\ldots,q-1$, is obtained by a right cyclic shift of the first row by $l$ positions.


For \ac{QC-LDPC} codes, each matrix $\mathbf{H}_{i,j}^{c}$ is a sparse
circulant matrix or a null matrix, and $\mathbf{H}$ is free of short
cycles in its associated Tanner graph.
A special family of \ac{QC-LDPC} codes is obtained by using circulant 
permutation matrices as $\mathbf{H}_{i,j}^{c}$ \cite{Fossorier2004}.
In this case, each circulant permutation
matrix is represented as a power of the unitary circulant permutation
matrix $\mathbf{P}$. The identity matrix is $\mathbf{P}^0$.
For this kind of \ac{QC-LDPC} codes, choosing a prime $q$ ensures some
desirable properties \cite{Fossorier2004}.
Array LDPC codes are based on circulant permutation matrices, and require a prime $q$.
Also the codes used in \cite{Tanner2004} are based on circulant permutation matrices
and often use a prime $q$ (although it is not mandatory).
Hence, from now on we consider only prime values of $q$.
Let $\mathbf{\Delta} = \{\Delta_0,\Delta_1,\Delta_2,\ldots ,\Delta_{r_0-1}\}$ be a set of $r_0$
distinct integers, with $\Delta_0<\Delta_1<\Delta_2<\cdots <\Delta_{r_0-1} < q$, $0 < r_0 < n_0\le q$,
and let us define
\begin{equation}
\mathbf{H}_a=\begin{bmatrix}
{0}&{\Delta_0}&{2\Delta_0}&
    \ldots &{(n_0-1)\Delta_0}\\
{0}&{\Delta_1}&{2\Delta_1}& 
    \ldots &{(n_0-1)\Delta_1}\\
{0}&{\Delta_2}&{2\Delta_2}& 
    \ldots &{(n_0-1)\Delta_2}\\
\vdots       &\vdots              & \vdots              &  \ddots & \vdots  \\
{0}&{\Delta_{r_0-1}}& {2\Delta_{r_0-1}} &
   \ldots &{(n_0-1)\Delta_{r_0-1}}
\end{bmatrix}.
\label{eq:Ha}
\end{equation}

An array \ac{LDPC} code is defined by the parity-check matrix $\mathbf{H}$ which
is obtained by using the elements of $\mathbf{H}_a$ as the exponents of $\mathbf{P}$.
Due to the primality of $q$, $\mathbf{H}$ is free of length-$4$ cycles.
If $n_0=q$, we have a \textit{full length} array \ac{LDPC} code.
In fact, for $n_0>q$, the code becomes trivial, and its minimum distance drops to $2$.
Choosing $n_0<q$ instead produces a \textit{shortened} array \ac{LDPC} code.
Reordering the columns of $\mathbf{H}$ yields an equivalent code. Reordering is usually performed by moving
blocks of $q$ columns each, such that the \ac{QC} form of $\mathbf{H}$ is preserved.
Shortening can also be performed after having reordered the columns of $\mathbf{H}$, thus
obtaining a code which is not necessarily equivalent to the non-reordered shortened array code.
Another classification of array codes depends on the choice of $\mathbf{\Delta}$:
if $\mathbf{\Delta} = \{0,1,2,\ldots ,{r_0-1}\}$, the code is called a \textit{proper} array LDPC code,
otherwise it is an \textit{improper} array LDPC code.

Theoretical arguments can be used to predict the minimum distance of array \ac{LDPC} codes,
or to find bounds on it.
The minimum distance of proper full length array \ac{LDPC} codes can be upper bounded by using the 
approach proposed in \cite{Yang2003, Sugiyama2008}, which also simplifies the search of low weight 
codewords.
The minimum distance of proper full length array \ac{LDPC} codes can be improved by resorting to improper \cite{Esmaeili2011}
and shortened codes \cite{Milenkovic2006}, thus approaching the values achievable through other design
techniques based on circulant permutation matrices. 
A general form for the generator matrix of array \ac{LDPC} codes has been presented in \cite{Baldi2012},
and helps deriving general upper bounds on their minimum distance.


\section{Array Convolutional LDPC codes \label{sec:ACLDPC} }

Starting from the parity-check matrix $\mathbf{H}$ of an array \ac{LDPC} code, we aim at
obtaining an \textit{unwrapped} version of it, to form the semi-infinite binary parity-check matrix
$\Hconv$ of an \ac{RTI-LDPCC} code.
For this purpose, let us rearrange the rows and the columns of $\mathbf{H}$ as follows:
permute the rows according to the ordering $0, q, 2q, \ldots , (r_0-1)q, 1, q+1, 2q+1, \ldots , (r_0-1)q+1, \ldots, q-1, 2q-1, 3q-1, \ldots, r_0q-1$
and then permute the columns according to the ordering $0, q, 2q, \ldots , (n_0-1)q, 1, q+1, 2q+1, \ldots , (n_0-1)q+1, \ldots, q-1, 2q-1, 3q-1, \ldots, n_0q-1$.
This way, we swap the inner and outer structures of the original matrix, which is a block of circulants, and obtain the parity-check matrix of an equivalent code in the form of a circulant of blocks,
that is

\begin{equation}
\mathbf{H}=\left[\begin{array}{llll}
\mathbf{H}_{0} & \mathbf{H}_{1} & \ldots & \mathbf{H}_{q-1}\\
\mathbf{H}_{q-1} & \mathbf{H}_{0} & \ldots & \mathbf{H}_{q-2}\\
\vdots & \vdots & \ddots & \vdots\\
\mathbf{H}_{1} & \mathbf{H}_{2} & \ldots & \mathbf{H}_{0}\end{array}\right],
\label{eq:HCircOfBlocks}
\end{equation}
where each block $\mathbf{H}_{i}$ is of size $r_{0}\times n_{0}$.
Starting from $\mathbf{H}$ in \eqref{eq:HCircOfBlocks}, we apply the unwrapping method in \cite{Felstrom1999}.
However, differently from \cite{Felstrom1999}, where time-varying convolutional 
codes are designed, we obtain time-invariant codes.
In fact, using the unwrapping in \cite{Felstrom1999} with a cutting pattern such 
that we repeatedly move $n_0$ positions to the right and then $r_0$ positions down, we obtain the
following semi-infinite parity-check matrix:

\begin{equation}
\Hconv = \left[\begin{array}{cccccc}
\HH_0 			& \0 					& \0 					& \cdots \\
\HH_{q-1}		& \HH_0				& \0 					& \cdots \\
\HH_{q-2}		& \HH_{q-1}		& \HH_0				& \cdots \\
\vdots 			& \vdots 			& \vdots 			& \ddots \\
\HH_1				& \HH_2				& \HH_3				& \ddots \\
\0 					& \HH_1				& \HH_2				& \ddots \\
\0 					& \0 					& \HH_1				& \ddots \\
\0 					& \0 					& \0					& \ddots \\
\vdots			& \vdots			& \vdots			& \ddots \\
\end{array}\right],
\label{eq:Hconv}
\end{equation}
which defines a time-invariant convolutional code with
syndrome former matrix $\HH_s = \left[ \HH_0^T | \HH_{q-1}^T | \HH_{q-2}^T | \ldots | \HH_1^T \right]$,
where $^T$ denotes transposition.
We say that this code is regular since $\HH_s$ is regular. Actually, the code defined by $\Hconv$ is column-wise regular
and only asymptotically row-wise regular, due to the effect of termination, which however is negligible for long block lengths
and sliding-window decoding. We have:
\begin{itemize}
\item Asymptotic code rate: $R = k_0/n_0$.
\item Parity-check matrix column weight: $r_0$.
\item Syndrome former memory order: $m_s = q$.
\item Syndrome former constraint length: $v_s = q \cdot n_0$.
\end{itemize}

\section{Comparison with previous solutions \label{sec:Comparison} }

An \ac{LDPCC} code is often obtained by designing an
\ac{LDPC} block code and then \textit{unwrapping} it. Two main methods exist for this purpose: the former starts from
a specific class of \ac{QC-LDPC} block codes to obtain time-invariant \ac{LDPCC} codes \cite{Tanner2004}, 
whereas the latter starts from generic \ac{LDPC} block codes to obtain time-varying \ac{LDPCC} codes \cite{Felstrom1999}.

The method we propose, described in Section \ref{sec:ACLDPC}, can be seen as a variant of them both,
with the main difference of using another class of \ac{QC-LDPC} block codes as the starting point.
In general, the codes designed according to \cite{Felstrom1999} are time-varying, while we obtain time invariant codes
by using array \ac{LDPC} codes as our starting point. 
Concerning the method in \cite{Tanner2004}, its starting point is instead a \ac{QC-LDPC} block code obtained from
the following exponent matrix:
\begin{equation}
\mathbf{H}_a^{[6]} = \begin{bmatrix}
1								& a 									& a^2 								& \ldots & a^{n_0-1} \\
b 							& a  b 								& a^2  b 							& \ldots & a^{n_0-1}  b \\
b^2							& a  b^2							& a^2  b^2						& \ldots & a^{n_0-1}  b^2 \\
\vdots					& \vdots 							& \vdots 							& \vdots & \ddots \\
b^{r_0-1} 			& a  b^{r_0-1} 			& a^2  b^{r_0-1}			& \ldots & a^{n_0-1}  b^{r_0-1} \end{bmatrix}
\label{eq:TannerExpMatrix}
\end{equation}
where $a$ and $b$ are two integers with multiplicative orders equal to $n_0$ and $r_0$
modulo $m$, respectively, and $m$ is an integer $> r_0 \cdot n_0$, such that $r_0$ and $n_0$ divide $m-1$.
Choosing the smallest $m$ with these characteristics allows to minimize the constraint length.
The elements of $\mathbf{H}_a^{[6]}$ (which are modulo $m$) are used as the exponents of the unitary
circulant permutation matrix $\mathbf{P}$ of size $q = m$ to create the parity-check matrix of a \ac{QC-LDPC} block code.
An \ac{RTI-LDPCC} code is obtained from the \ac{QC-LDPC} block code
through the unwrapping procedure described in \cite{Tanner2004}.
The memory order of the resulting code equals the maximum value taken
by the difference between the largest and the smallest elements in each row of $\mathbf{H}_a^{[6]}$, increased by one \cite{Pusane2011}.
As acknowledged in \cite{Tanner2004}, these \ac{LDPCC} codes typically have large constraint lengths.

We observe that the exponent matrix \eqref{eq:TannerExpMatrix} defines the same code that can be obtained 
by starting from an array \ac{LDPC} code with $\mathbf{\Delta} = \{1,b,b^2,\ldots ,b^{r_0-1}\}$,
and then performing column reordering, followed by shortening.
So, the codes used as the starting point in \cite{Tanner2004} 
are a special case of array \ac{LDPC} codes.
Inspired by this observation, we use more general array \ac{LDPC} codes than those obtained
from \eqref{eq:TannerExpMatrix} as our starting point to design \ac{RTI-LDPCC} codes.
After fixing $\mathbf{\Delta}$, we use the matrix \eqref{eq:Ha} as
the starting point of the unwrapping procedure, and this allows to obtain unwrapped codes with smaller constraint 
length than when starting from the matrix \eqref{eq:TannerExpMatrix}, as in \cite{Tanner2004}.
In addition, at high code rates, the performance of the unwrapped codes and, in some cases, their minimum distance 
are improved, as we will show in Section \ref{sec:Examples}.
Moreover, this allows to find a tradeoff between the constraint length and the performance by varying $q$, as we will also show next.

Concerning the unwrapping method, we use the one described in Section \ref{sec:ACLDPC}, which exploits the 
technique proposed in \cite{Felstrom1999}.
Alternatively, we could replace the matrix \eqref{eq:TannerExpMatrix} with \eqref{eq:Ha} and use again the 
unwrapping method in \cite{Tanner2004}. This would produce similar results, as we will show in Example \ref{exa:Rate25}.
We prefer to use the procedure in Section \ref{sec:ACLDPC} because it makes easier to outline the 
structure of the syndrome former matrix $\HH_s$.

\begin{Exa}
\label{exa:Rate25}
In order to provide an explicit simple example of the proposed method,
let us consider the following two exponent matrices defining, respectively, a
full length proper array \ac{LDPC} code with $r_0=3, n_0=q=5$ and a 
shortened proper array \ac{LDPC} code with $r_0=3, n_0=5, q=7$:

\begin{equation}
\HH'_a=\begin{bmatrix}
0 & 0 & 0 & 0 & 0 \\
0 & 1 & 2 & 3 & 4 \\
0 & 2 & 4 & 1 & 3 \end{bmatrix}, \ \ 
\HH''_a=\begin{bmatrix}
0 & 0 & 0 & 0 & 0 \\
0 & 1 & 2 & 3 & 4 \\
0 & 2 & 4 & 6 & 1 \end{bmatrix}.
\label{eq:HexpArray}
\end{equation}

We can unwrap them to obtain two \ac{RTI-LDPCC} codes with rate $2/5$.
By using the method in Section \ref{sec:ACLDPC}, we get two
\ac{AC-LDPC} codes having the syndrome former matrices reported in Figs. \ref{fig:H0t} (a) and (c),
respectively.
These two \ac{LDPCC} codes have $(m_s,v_s)$ equal to $(5,25)$ and $(7,35)$. 
If we instead follow the approach in \cite{Tanner2004}, starting from the matrix \eqref{eq:TannerExpMatrix}, 
we are forced to accept longer constraint lengths.
An \ac{LDPCC} code with the same parameters (see \cite[Example 6]{Tanner2004})
would have $(m_s=22,v_s=105)$.
Finally, we observe that, by using the unwrapping technique in \cite{Tanner2004}, but starting from
$\HH'_a$ and $\HH''_a$ in \eqref{eq:HexpArray}, we obtain the syndrome former matrices reported in Figs. \ref{fig:H0t} (b) and (d).
As anticipated, these are very similar to those obtained through the alternative unwrapping technique
we consider, since they are row-reordered versions of them.

\begin{figure}
\centering
\begin{tabular}{@{\hspace{-1mm}}c@{\hspace{-1mm}}@{\hspace{-1mm}}c@{\hspace{-1mm}}@{\hspace{-1mm}}c@{\hspace{-1mm}}@{\hspace{-1mm}}c@{\hspace{-1mm}}}
\subfloat[]{
$
\scriptsize
\setlength{\arraycolsep}{3pt}
\begin{bmatrix}
1 & 1 & 1 & 1 & 1 \\
1 & 0 & 0 & 0 & 0 \\
1 & 0 & 0 & 0 & 0 \\
0 & 0 & 0 & 0 & 0 \\
0 & 0 & 0 & 0 & 1 \\
0 & 0 & 1 & 0 & 0 \\
0 & 0 & 0 & 0 & 0 \\
0 & 0 & 0 & 1 & 0 \\
0 & 0 & 0 & 0 & 1 \\
0 & 0 & 0 & 0 & 0 \\
0 & 0 & 1 & 0 & 0 \\
0 & 1 & 0 & 0 & 0 \\
0 & 0 & 0 & 0 & 0 \\
0 & 1 & 0 & 0 & 0 \\
0 & 0 & 0 & 1 & 0 \end{bmatrix}
$
} &
\subfloat[]{
$
\scriptsize
\setlength{\arraycolsep}{3pt}
\begin{bmatrix}
0 & 0 & 1 & 0 & 0 \\
0 & 0 & 0 & 0 & 1 \\
0 & 0 & 0 & 0 & 0 \\
0 & 0 & 0 & 0 & 1 \\
0 & 0 & 0 & 1 & 0 \\
0 & 0 & 0 & 0 & 0 \\
0 & 1 & 0 & 0 & 0 \\
0 & 0 & 1 & 0 & 0 \\
0 & 0 & 0 & 0 & 0 \\
0 & 0 & 0 & 1 & 0 \\
0 & 1 & 0 & 0 & 0 \\
0 & 0 & 0 & 0 & 0 \\
1 & 0 & 0 & 0 & 0 \\
1 & 0 & 0 & 0 & 0 \\
1 & 1 & 1 & 1 & 1 \end{bmatrix}
$
} &
\subfloat[]{
$
\scriptsize
\setlength{\arraycolsep}{3pt}
\begin{bmatrix}
1 & 1 & 1 & 1 & 1 \\
1 & 0 & 0 & 0 & 0 \\
1 & 0 & 0 & 0 & 0 \\
0 & 0 & 0 & 0 & 0 \\
0 & 0 & 0 & 0 & 0 \\
0 & 0 & 0 & 1 & 0 \\
0 & 0 & 0 & 0 & 0 \\
0 & 0 & 0 & 0 & 0 \\
0 & 0 & 0 & 0 & 0 \\
0 & 0 & 0 & 0 & 0 \\
0 & 0 & 0 & 0 & 1 \\
0 & 0 & 1 & 0 & 0 \\
0 & 0 & 0 & 0 & 0 \\
0 & 0 & 0 & 1 & 0 \\
0 & 0 & 0 & 0 & 0 \\
0 & 0 & 0 & 0 & 0 \\
0 & 0 & 1 & 0 & 0 \\
0 & 1 & 0 & 0 & 0 \\
0 & 0 & 0 & 0 & 0 \\
0 & 1 & 0 & 0 & 0 \\
0 & 0 & 0 & 0 & 1 \end{bmatrix}
$
}
&
\subfloat[]{
$
\scriptsize
\setlength{\arraycolsep}{3pt}
\begin{bmatrix}
0 & 0 & 0 & 1 & 0 \\
0 & 0 & 0 & 0 & 0 \\
0 & 0 & 0 & 0 & 0 \\
0 & 0 & 0 & 0 & 0 \\
0 & 0 & 0 & 0 & 0 \\
0 & 0 & 0 & 0 & 0 \\
0 & 0 & 1 & 0 & 0 \\
0 & 0 & 0 & 0 & 1 \\
0 & 0 & 0 & 0 & 0 \\
0 & 0 & 0 & 0 & 0 \\
0 & 0 & 0 & 1 & 0 \\
0 & 0 & 0 & 0 & 0 \\
0 & 1 & 0 & 0 & 0 \\
0 & 0 & 1 & 0 & 0 \\
0 & 0 & 0 & 0 & 0 \\
0 & 0 & 0 & 0 & 1 \\
0 & 1 & 0 & 0 & 0 \\
0 & 0 & 0 & 0 & 0 \\
1 & 0 & 0 & 0 & 0 \\
1 & 0 & 0 & 0 & 0 \\
1 & 1 & 1 & 1 & 1 \end{bmatrix}
$
}
\end{tabular}
\caption{Syndrome former matrices of \ac{LDPC} convolutional codes obtained from the codes defined by $\HH'_a$ (a), (b) and $\HH''_a$ (c), (d) in \eqref{eq:HexpArray}, through the use of the unwrapping technique in Section \ref{sec:ACLDPC} (a), (c) and that in \cite{Tanner2004} (b), (d).}
\label{fig:H0t}
\end{figure}
\end{Exa}

Another benchmark for the proposed class of codes are \ac{PDC-LDPC} codes \cite{Baldi2012b}.
They are characterized by a very small syndrome former constraint length,
and may have fixed minimum distance, independently of the code rate.
However, the very small constraint length of \ac{PDC-LDPC} codes is paid
in terms of performance, which barely approaches that of the codes designed
following \cite{Tanner2004}.
As we will show in Section \ref{sec:Examples}, the newly proposed \ac{AC-LDPC} codes 
instead allow to trade the syndrome former constraint length for performance, thus
outperforming both these previous approaches.

\section{Performance assessment \label{sec:Examples} }

We consider some code examples and simulate coded
transmissions over the additive white Gaussian noise channel, with binary phase shift keying.
\ac{LDPC} decoding is performed through the sum-product algorithm with log-likelihood 
ratios, with $100$ maximum iterations, working over blocks of $60000$ bits.
These choices ensure that the decoding algorithm achieves optimal performance for all
the considered codes.

\begin{figure}[!t]
\centering
\subfloat[]{\includegraphics[keepaspectratio,width=44mm]{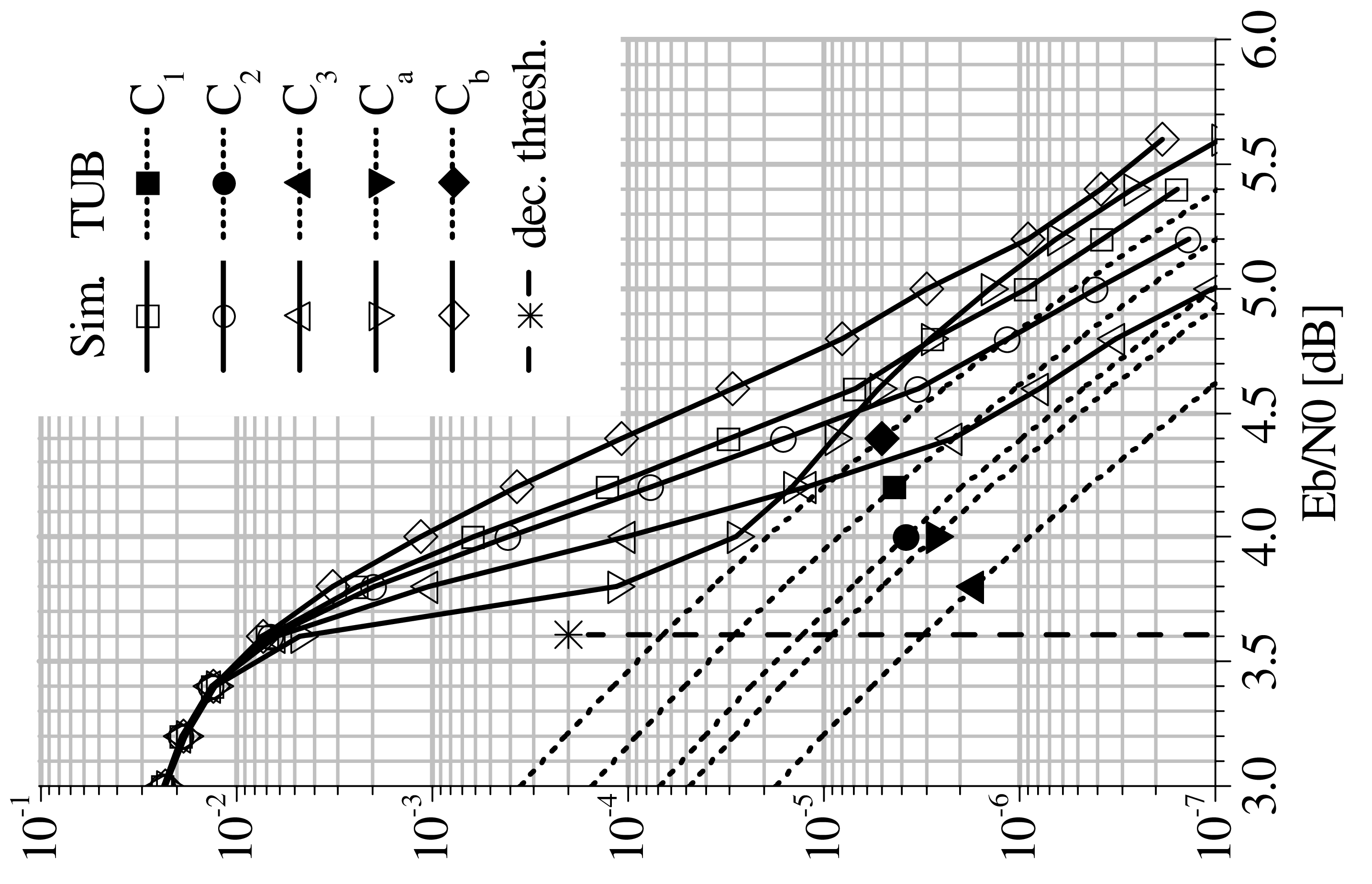}}
\subfloat[]{\includegraphics[keepaspectratio,width=44mm]{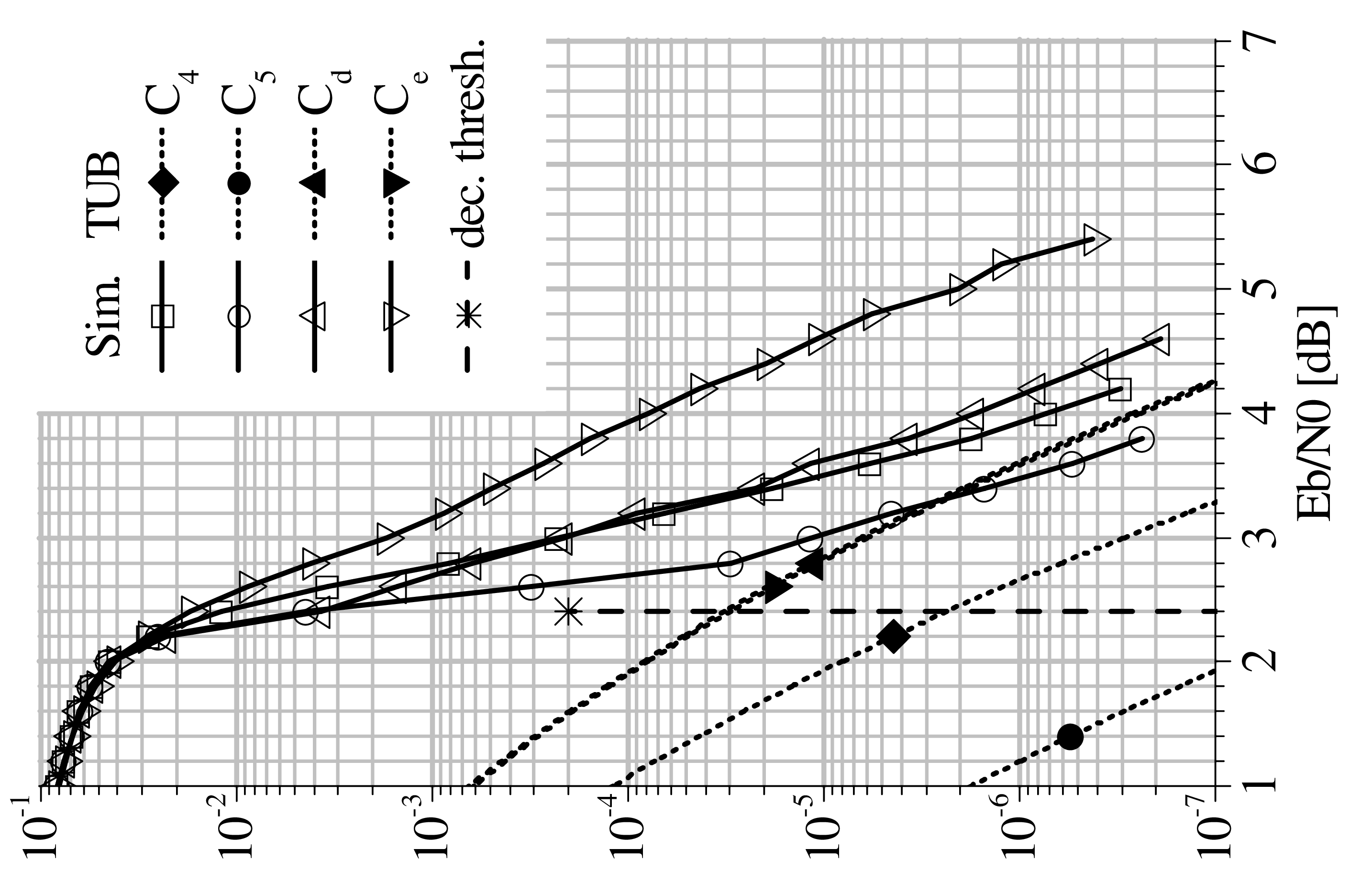}}
\caption{Simulated \ac{BER} and \ac{TUB} curves and decoding thresholds for the codes in Tables \ref{tab:ACLDPC} and \ref{tab:OtherCodes} with (a) rate $0.9$ and (b) rate $0.75$.}
\label{fig:Sim}
\end{figure}

The code parameters are summarized in Tables \ref{tab:ACLDPC} and \ref{tab:OtherCodes},
where $R$ denotes the code rate, $w$ is the parity-check matrix column weight and $d$ the minimum distance.
The latter has been estimated through Montecarlo simulations, by isolating the unique error vectors generating
the low weight codewords observed. This way, as done in \cite{Snow2007}, we have estimated the 
spectrum of the lowest weights for each code, and this has been used to compute the corresponding 
\ac{TUB} \cite{Poltyrev1994}.
Table \ref{tab:ACLDPC} reports the details of the shortened proper and improper \ac{AC-LDPC} codes we
have designed, whereas Table \ref{tab:OtherCodes} provides the parameters of the other codes we have used
as a benchmark, designed according to \cite{Tanner2004} and \cite{Baldi2012b}.
The weight spectra are not reported for the lack of space.
For improper \ac{AC-LDPC} codes, $\mathbf{\Delta}$ has been optimized heuristically.
Fig. \ref{fig:Sim} shows the \ac{BER} and \ac{TUB} curves for all codes, grouped by code rate.
The decoding thresholds obtained through density evolution are also plotted.

\begin{table}[!t]
\renewcommand{\arraystretch}{1.1}
\caption{Examples of \ac{AC-LDPC} codes}
\label{tab:ACLDPC}
\centering
\begin{tabular}{|c|c|c|c|c|c|c|c|c|}
\hline
Code & $R$ & $r_0$ & $n_0$ & $w$ & $\mathbf{\Delta}$ &  $q$ & $v_s$ & $d$\\
\hline
$C_1$ & $0.9$ & $3$ & $30$ & $3$ & $\{0,1,2\}$ & $43$ & $1290$ & $6$ \\
\hline
$C_2$ & $0.9$ & $3$ & $30$ & $3$ & $\{0,11,37\}$ & $43$ & $1290$ & $6$ \\
\hline
$C_3$ & $0.9$ & $3$ & $30$ & $3$ & $\{0,11,37\}$ & $71$ & $2130$ & $6$ \\
\hline
$C_4$ & $0.75$ & $4$ & $16$ & $4$ & $\{0,1,2,3\}$ & $71$ & $1136$ & $10$ \\
\hline
$C_5$ & $0.75$ & $4$ & $16$ & $4$ & $\{0,11,37,70\}$ & $71$ & $1136$ & $12$ \\
\hline
\end{tabular}
\end{table}

\begin{table}[!t]
\renewcommand{\arraystretch}{1.1}
\caption{Examples of other regular time invariant \ac{LDPC} convolutional codes}
\label{tab:OtherCodes}
\centering
\begin{tabular}{|c|c|c|c|c|c|c|c|}
\hline
Code & $R$ & $r_0$ & $n_0$ & $w$ & $\{m, a, b\}$ & $v_s$ & $d$  \\
\hline
$C_a$ \cite{Tanner2004} & $0.9$ & $3$ & $30$ & $3$ & $\{151, 23, 32\}$ & $4500$ & $6$ \\
\hline
$C_b$ \cite{Baldi2012b} & $0.9$ & $1$ & $10$ & $3$ & $-$ & $400$ & $6$ \\
\hline
$C_d$ \cite{Tanner2004} & $0.75$ & $4$ & $16$ & $4$ & $\{97, 8, 22\}$ & $1536$ & $8$ \\
\hline
$C_e$ \cite{Baldi2012b} & $0.75$ & $1$ & $4$ & $4$ & $-$ & $152$ & $8$ \\
\hline
\end{tabular}
\end{table}

From the tables we observe that all the codes with rate $0.9$ have minimum distance $6$.
However, the code $C_3$ has the lowest multiplicity of minimum weight codewords, which reflects into a better \ac{TUB}.
This is also confirmed by the simulation curves, since $C_3$ achieves a gain of $0.6$ dB or more over $C_a$ and $C_b$.
In addition, it has a constraint length $v_s$ more than halved than that of $C_a$.
The codes $C_1$ and $C_2$ achieve a further reduction in the constraint length, but at the cost of some loss in performance.
The benefits of the proposed design technique are even more evident for codes with rate $0.75$, 
at which \ac{AC-LDPC} codes also achieve higher minimum distances than those 
designed according to \cite{Tanner2004} and \cite{Baldi2012b}.
In fact, $C_4$ and $C_5$ have better \ac{BER} and \ac{TUB} curves than $C_d$ and $C_e$, and also achieve a
reduction in the constraint length with respect to $C_d$.
The code $C_e$ has the worst performance, but its constraint length is largely below that of all the other codes.
As expected, these codes incur some performance loss with respect to irregular time-varying \ac{LDPCC} codes with
optimized degree distributions. For example, the rate $3/4$ code reported in \cite[Fig. 9]{Pusane2011} exhibits a gain
of about $0.8$ dB over $C_5$.
However, it has a constraint length more than doubled, and requires to deal with an irregular and time-varying structure.

\end{document}